\documentclass[conference]{IEEEtran}
\usepackage{cite}
\usepackage{amsmath,amssymb,amsfonts}
\usepackage{graphicx}
\usepackage{textcomp}
\usepackage{xcolor}
\usepackage{booktabs}
\usepackage{tikz}
\usetikzlibrary{positioning,fit,arrows.meta}
\usepackage[hidelinks]{hyperref}
\hypersetup{pdftitle={Scaffold, Fill, Verify: An Industrial Experience Report on Agentic Generation of Integration Pipelines},pdfauthor={Luciano Lucindo Chaves, Renan Novais, Pablo Calvache}}

\newcommand{\clusterWidthMax}{1.32}
\newcommand{\clusterWidthMin}{0.98}
\newcommand{\cmpAAllCost}{$-$49.3\%}
\newcommand{\cmpAAllCostAbs}{49.3\%}
\newcommand{\cmpAAllCostCI}{[$-$55.2; $-$42.1]}
\newcommand{\cmpAAllCps}{$-$55.3\%}
\newcommand{\cmpAAllCpsCI}{[$-$61.8; $-$47.6]}
\newcommand{\cmpAAllJudge}{+0.057}
\newcommand{\cmpAAllJudgeCI}{[0.019; 0.095]}
\newcommand{\cmpAAllPassCrit}{+17.6\,pp}
\newcommand{\cmpAAllPassCritCI}{[4.5; 31.4]}
\newcommand{\cmpAAllPassRaw}{+4.3\,pp}
\newcommand{\cmpAAllPassRawCI}{[$-$8.6; 17.9]}
\newcommand{\cmpAAllPassSym}{+11.9\,pp}
\newcommand{\cmpAAllPassSymCI}{[1.0; 24.0]}
\newcommand{\cmpAAllRules}{$-$0.033}
\newcommand{\cmpAAllRulesCI}{[$-$0.080; 0.014]}
\newcommand{\cmpAAllTime}{$-$41.0\%}
\newcommand{\cmpAAllTimeAbs}{41.0\%}
\newcommand{\cmpAAllTimeCI}{[$-$45.6; $-$35.1]}
\newcommand{\cmpADonePassSym}{+19.8\,pp}
\newcommand{\cmpADonePassSymCI}{[4.8; 36.5]}
\newcommand{\cmpADtwoPassSym}{0.0\,pp}
\newcommand{\cmpADtwoPassSymCI}{[$-$14.3; 16.1]}
\newcommand{\cmpBAllJudge}{+0.050}
\newcommand{\cmpBAllJudgeCI}{[0.019; 0.082]}
\newcommand{\cmpBAllPassSym}{$-$1.6\,pp}
\newcommand{\cmpBAllPassSymCI}{[$-$4.7; 0.0]}
\newcommand{\cmpBAllPassSymCommon}{+10.9\,pp}
\newcommand{\cmpBAllPassSymCommonCI}{[2.3; 20.3]}
\newcommand{\cmpBAllPassSymCommonNM}{+4.7\,pp}
\newcommand{\cmpBAllPassSymCommonNMCI}{[$-$0.8; 11.7]}
\newcommand{\cmpBAllPassSymRawOff}{$-$9.4\,pp}
\newcommand{\cmpBAllPassSymRawOffCI}{[$-$18.8; $-$1.6]}
\newcommand{\cmpBAllRules}{$-$0.027}
\newcommand{\cmpBAllRulesCI}{[$-$0.068; 0.015]}
\newcommand{\cmpBAllTime}{$-$63.7\%}
\newcommand{\cmpBAllTimeCI}{[$-$67.9; $-$58.9]}
\newcommand{\cmpBDevCost}{$-$74.2\%}
\newcommand{\cmpBDevCostAbs}{74.2\%}
\newcommand{\cmpBDevCostCI}{[$-$76.7; $-$71.4]}
\newcommand{\cmpBDevJudge}{+0.041}
\newcommand{\cmpBDevJudgeCI}{[0.006; 0.076]}
\newcommand{\cmpBDevPassSym}{$-$1.0\,pp}
\newcommand{\cmpBDevPassSymCI}{[$-$8.6; 6.7]}
\newcommand{\cmpBDevPassSymCommon}{+2.4\,pp}
\newcommand{\cmpBDevPassSymCommonCI}{[$-$4.8; 10.5]}
\newcommand{\cmpBDevPassSymRawOff}{$-$3.8\,pp}
\newcommand{\cmpBDevPassSymRawOffCI}{[$-$14.3; 5.7]}
\newcommand{\cmpBDevRules}{+0.012}
\newcommand{\cmpBDevRulesCI}{[$-$0.033; 0.060]}
\newcommand{\cmpBDevTime}{$-$61.9\%}
\newcommand{\cmpBDevTimeAbs}{61.9\%}
\newcommand{\cmpBDevTimeCI}{[$-$65.8; $-$58.0]}
\newcommand{\cmpCriticDtwoCostAbs}{34.4\%}
\newcommand{\cmpCriticDtwoJudge}{$-$0.058}
\newcommand{\cmpCriticDtwoJudgeCI}{[$-$0.101; $-$0.010]}
\newcommand{\cmpCriticDtwoPassSym}{$-$2.4\,pp}
\newcommand{\cmpCriticDtwoPassSymCI}{[$-$14.9; 8.9]}
\newcommand{\cmpCriticDtwoRules}{$-$0.149}
\newcommand{\cmpCriticDtwoRulesCI}{[$-$0.213; $-$0.080]}
\newcommand{\cmpCriticDtwoTimeAbs}{27.5\%}
\newcommand{\cmpOldDoneCostAbs}{72.2\%}
\newcommand{\cmpOldDoneTimeAbs}{70.4\%}
\newcommand{\condPrBDoneCostIdx}{1.84}
\newcommand{\condPrBDoneJudge}{0.624}
\newcommand{\condPrBDonePassSym}{96.8}
\newcommand{\condPrBDoneRuns}{3}
\newcommand{\condPrBDoneTime}{1037}
\newcommand{\condPrBDtwoCostIdx}{1.93}
\newcommand{\condPrBDtwoJudge}{0.659}
\newcommand{\condPrBDtwoPassSym}{95.2}
\newcommand{\condPrBDtwoTime}{1083}
\newcommand{\condPrBHrealJudge}{0.602}
\newcommand{\condPrBHrealPassSym}{100.0}
\newcommand{\condPrBHrealRuns}{4}
\newcommand{\condPrBHrealTime}{1224}
\newcommand{\condPrBHsynJudge}{0.703}
\newcommand{\condPrBHsynPassSym}{100.0}
\newcommand{\condPrBHsynRuns}{4}
\newcommand{\condPrBHsynTime}{903}
\newcommand{\condPrDoneCostIdx}{1.00}
\newcommand{\condPrDoneJudge}{0.628}
\newcommand{\condPrDonePassRaw}{73.8}
\newcommand{\condPrDonePassSym}{78.6}
\newcommand{\condPrDoneRuns}{2}
\newcommand{\condPrDoneTime}{620}
\newcommand{\condPrDtwoCostIdx}{1.24}
\newcommand{\condPrDtwoJudge}{0.629}
\newcommand{\condPrDtwoPassSym}{92.9}
\newcommand{\condPrDtwoRuns}{2}
\newcommand{\condPrDtwoTime}{685}
\newcommand{\condSkBDoneCostIdx}{0.43}
\newcommand{\condSkBDoneJudge}{0.678}
\newcommand{\condSkBDonePassSym}{96.8}
\newcommand{\condSkBDoneRuns}{3}
\newcommand{\condSkBDoneTime}{369}
\newcommand{\condSkBDtwoCostIdx}{0.57}
\newcommand{\condSkBDtwoJudge}{0.680}
\newcommand{\condSkBDtwoPassSym}{92.9}
\newcommand{\condSkBDtwoTime}{453}
\newcommand{\condSkBHrealJudge}{0.667}
\newcommand{\condSkBHrealPassSym}{97.2}
\newcommand{\condSkBHrealRuns}{4}
\newcommand{\condSkBHrealTime}{492}
\newcommand{\condSkBHsynJudge}{0.735}
\newcommand{\condSkBHsynPassSym}{100.0}
\newcommand{\condSkBHsynRuns}{4}
\newcommand{\condSkBHsynTime}{283}
\newcommand{\condSkDoneCostIdx}{0.50}
\newcommand{\condSkDoneJudge}{0.698}
\newcommand{\condSkDonePassCrit}{98.4}
\newcommand{\condSkDonePassRaw}{81.0}
\newcommand{\condSkDonePassSym}{98.4}
\newcommand{\condSkDoneRuns}{3}
\newcommand{\condSkDoneTime}{349}
\newcommand{\condSkDtwoCostIdx}{0.65}
\newcommand{\condSkDtwoJudge}{0.667}
\newcommand{\condSkDtwoPassSym}{92.9}
\newcommand{\condSkDtwoRuns}{4}
\newcommand{\condSkDtwoTime}{433}
\newcommand{\connBAll}{39}
\newcommand{\connBCases}{67}
\newcommand{\costPrBEvCases}{105}
\newcommand{\costSkBEvCases}{141}
\newcommand{\ctFullPrFail}{19}
\newcommand{\ctFullPrFailMock}{8}
\newcommand{\ctFullPrN}{219}
\newcommand{\ctFullPrPass}{200}
\newcommand{\ctFullSkN}{226}
\newcommand{\estBHrealCostRed}{72\%}
\newcommand{\estBHsynCostRed}{85\%}
\newcommand{\estPrBHrealCostIdx}{2.04}
\newcommand{\estPrBHsynCostIdx}{1.66}
\newcommand{\estPrBLoroErr}{9.5}
\newcommand{\estSkBHrealCostIdx}{0.57}
\newcommand{\estSkBHsynCostIdx}{0.26}
\newcommand{\estSkBLoroErr}{5.0}
\newcommand{\failAPrBuilds}{65}
\newcommand{\failAPrFails}{13}
\newcommand{\failASkBuilds}{119}
\newcommand{\failASkFails}{5}
\newcommand{\failASkInvalid}{2}
\newcommand{\failASkTest}{2}
\newcommand{\judgeBCred}{+0.179}
\newcommand{\judgeBCredCI}{[0.104; 0.252]}
\newcommand{\judgeBErr}{$-$0.013}
\newcommand{\judgeBErrCI}{[$-$0.053; 0.027]}
\newcommand{\prDoneCostGrowth}{1.8}
\newcommand{\prDoneSmallCallsA}{10}
\newcommand{\prDoneSmallCallsB}{29}
\newcommand{\prDoneTimeGrowth}{1.7}
\newcommand{\prDtwoCostGrowth}{1.5}
\newcommand{\prDtwoTimeGrowth}{1.6}
\newcommand{\primBetaBinP}{0.015}
\newcommand{\primDoneN}{21}
\newcommand{\primDonePos}{7}
\newcommand{\primGlmmP}{0.010}
\newcommand{\primProjCI}{[2.1; 23.6]}
\newcommand{\primSignFlipP}{0.040}
\newcommand{\refusA}{3}
\newcommand{\refusAll}{9}
\newcommand{\refusB}{6}
\newcommand{\replSkDonePassCrit}{93.7}
\newcommand{\revalBArtifacts}{466}
\newcommand{\setDoneActs}{15 [12--22]}
\newcommand{\setDoneDesign}{10 [8--12]}
\newcommand{\setDoneSpec}{9 [8--14]}
\newcommand{\setDoneStepsPr}{42 [34--56]}
\newcommand{\setDoneStepsSk}{56 [43--75]}
\newcommand{\setDtwoActs}{25 [19--36]}
\newcommand{\setDtwoDesign}{17 [11--22]}
\newcommand{\setDtwoProjects}{6}
\newcommand{\setDtwoSpec}{16 [10--23]}
\newcommand{\setDtwoStepsPr}{58 [38--84]}
\newcommand{\setDtwoStepsSk}{65 [54--95]}
\newcommand{\setExpansion}{2.9}
\newcommand{\setHrealActs}{27 [20--34]}
\newcommand{\setHrealDesign}{18 [14--23]}
\newcommand{\setHrealProjects}{9}
\newcommand{\setHrealSpec}{12 [9--15]}
\newcommand{\setHrealStepsPr}{62 [44--71]}
\newcommand{\setHrealStepsSk}{61 [49--73]}
\newcommand{\setHsynActs}{11 [8--14]}
\newcommand{\setHsynDesign}{9 [7--10]}
\newcommand{\setHsynSpec}{8 [6--9]}
\newcommand{\setHsynStepsPr}{27 [23--34]}
\newcommand{\setHsynStepsSk}{35 [27--43]}
\newcommand{\setPoolCases}{67}
\newcommand{\setRhoActs}{0.85}
\newcommand{\setRhoSpec}{0.34}
\newcommand{\skOwnStrictB}{26}
\newcommand{\triBExec}{21}
\newcommand{\triBHoCases}{3}
\newcommand{\triBHoN}{12}
\newcommand{\triBN}{22}
\newcommand{\triBPrHo}{1}
\newcommand{\triBSkHo}{11}

\newcommand{\code}[1]{\texttt{#1}}
\newcommand{\Done}{D\textsubscript{1}}
\newcommand{\Dtwo}{D\textsubscript{2}}
\newcommand{\Hreal}{H\textsubscript{cust}}
\newcommand{\Hsyn}{H\textsubscript{syn}}

\begin{document}
\bstctlcite{IEEEexample:BSTcontrol}

\title{Scaffold, Fill, Verify: An Industrial Experience Report on Agentic Generation of Integration Pipelines}

\author{\IEEEauthorblockN{Luciano Lucindo Chaves, Renan Novais and Pablo Calvache}
\IEEEauthorblockA{\textit{AI Engineering Team, Digibee}\\
S\~{a}o Paulo, Brazil\\
\{luciano.chaves, renan.novais, pablo.calvache\}@digibee.com}}

\maketitle

\begin{abstract}
A common design for agentic generators of executable artifacts lets one agent write the whole artifact and then loop over validation, testing and repair.
We report on replacing that design in the production pipeline builder of an enterprise integration platform, whose pipelines are written in a proprietary JSON-based domain-specific language. In the scaffolded builder that replaced it, one strong-model call plans the control-flow skeleton, and deterministic passes fix its shape. Small-model calls then fill each step in parallel. The platform validator and a mocked execution in the real engine drive a correction loop that routes each error to the step that caused it.
We compared the two builders in two campaigns three weeks apart, with one acceptance criterion for both.
In both campaigns the scaffolded builder cost less and built faster; on cases used during development its cost per case was \cmpAAllCostAbs{} lower in the first campaign and \cmpBDevCostAbs{} lower in the second, and its build time \cmpAAllTimeAbs{} and \cmpBDevTimeAbs{} shorter.
It also passed more often in the first campaign (once validator false positives were discounted for both builders); by the second, with both builders near the ceiling, the gap was no longer detectable and the production builder's cost had grown, and on 32 held-out cases we found no difference that did not depend on the measuring instrument.
We describe the architecture, a measurement protocol for builders with different internals, and lessons on validator false positives and on benchmarking against a baseline that changes during the study.
The scaffolded builder has since replaced the production builder.
\end{abstract}


\section{Introduction}

Integration platforms (iPaaS) let customers connect business systems with pipelines assembled from connectors.
Writing such a pipeline from a requirements document is skilled, repetitive work, which makes it an obvious target for LLM agents.
We work on one such platform, whose product included an agentic pipeline builder in production use throughout the study.
That builder is the baseline of this study.
It follows the generate-then-repair recipe of code generation~\cite{chen2024selfdebug}, run by an agent with tools: it generates the complete pipeline definition and then iterates over the platform validator and a test run until both succeed, a token budget is spent or a cycle limit is reached.

This recipe has three drawbacks.
Generation calls carry the whole artifact and much supporting information, and in the repair loop the context grows even with message windowing, so cost tends to grow with pipeline size.
Repair is global: a validator message about one step can cause the agent to rewrite others.
Stopping depends on exit conditions: if one fails to trigger, the loop keeps making calls after the work is done until a budget runs out; self-repair gains are often modest once their cost is counted~\cite{olausson2024selfrepair}.

We built a second builder around a different division of labour, which we call \emph{scaffold, fill, verify}.
We delegate control-flow design and connector selection to a strong model, while deterministic code enforces the structural schema of the artifact.
A small model fills each step's parameters with local context, in parallel.
The platform's validator and an instrumented execution in the real engine then drive a correction loop in which each error is routed back to the step that caused it.

Splitting generation into an outline and a fill has been published for enterprise low-code workflows~\cite{ayala2025lowcode,liu2025workteam,gschwind2025cag} and for parallel text generation~\cite{ning2024sot}.
The industrial studies evaluate by similarity to reference workflows or structural error rates, without execution.
Our contributions are:

\begin{itemize}
\item \textbf{An execution-grounded architecture} for a proprietary JSON-based domain-specific language (DSL), in which every LLM step after the fill is guarded by a monotonic acceptance rule, and validator messages are triaged into genuine errors, transient failures, provisioning gaps and false positives before any repair (Section~\ref{sec:arch}).
\item \textbf{A measurement protocol} that makes two builders with different internals comparable: one acceptance criterion over all cases, offline revalidation of both builders' artifacts with the same triage, a cross-test of artifacts the baseline never tested, and paired bootstrap intervals (Section~\ref{sec:design}).
\item \textbf{Evidence from two campaigns} against the production builder, on development and held-out cases. The acceptance advantage measured in October did not reappear in mid-November, on held-out cases or on reruns of the development cases, after the baseline had changed, while the efficiency advantage grew (Section~\ref{sec:results}).
\item \textbf{Lessons} for teams that build or evaluate such generators (Section~\ref{sec:lessons}).
\end{itemize}

\section{Context}
\label{sec:context}

\subsection{Pipelines and their definition language}

The study was conducted on a commercial, cloud-native iPaaS that serves more than 250 corporate customers in financial services, retail, logistics and healthcare, runs thousands of pipelines a day and processes billions of transactions a month.
A pipeline is written in a DSL whose concrete syntax is JSON.
It contains \emph{tracks}, each a sequence of \emph{steps}.
A step is a connector instance (REST call, database query, file transfer, JSON transformation, choice, for-each, do-while, sub-flow execution and others) with parameters.
Parameter values use an expression language with double-brace references to the current message, to earlier steps and to platform global variables, such as \code{\{\{ message.id \}\}}, \code{\{\{ step.alias.total \}\}} or \code{\{\{ global.name \}\}}, and functions for conversion, dates and conditionals.
Each connector has its own parameter schema.
Error handling is expressed as separate tracks attached to steps.

The platform exposes two services that both builders use: a \emph{validator}, which checks a definition against connector schemas and the expression grammar, and a \emph{test engine}, which executes a pipeline.
The language is proprietary, so no model has seen meaningful amounts of it in training; this is the low-resource, domain-specific setting surveyed in~\cite{joel2024lowresource}.
A definition that is well-formed JSON can still fail in several ways: a parameter the connector's schema does not accept, a reference to a step that has not run or to a field it does not produce, a payload whose shape the next step does not expect, or a construct that the validator accepts and the engine rejects.

\subsection{Input}

A build request contains a requirements specification in prose (systems, authentication, entities, field mappings, business rules, error handling) and a \emph{workflow design}: a structured outline of activities, branches, loops and parallel sections that acts as the control-flow blueprint.
Both builders receive this activity-level flow as input and map it to concrete steps, connectors and data-flow expressions.
A catalogue of connectors and reusable building-block patterns is available through tool calls.

\subsection{Baseline: the production builder}

The production builder has two parts.
A tool-using agent on the strong model writes the complete definition, using the same connector catalogue and building-block patterns.
An incremental semi-deterministic validation procedure, with its own agent on the small model, checks the definition against the validator and repairs it.
Once the definition is valid, the builder runs a test with mocked external services and repairs again.
It does not test a definition that fails validation.
Two of the authors built the production builder, and all three have developed it since; it remained under active development during the study, and we return to the consequences in Section~\ref{sec:lessons}.

\section{The scaffolded builder}
\label{sec:arch}

\begin{figure}[t]
\centering
\begin{tikzpicture}[
  font=\footnotesize, node distance=1.4mm,
  box/.style={draw, rounded corners=1pt, minimum height=4.6mm, text width=0.80\columnwidth, align=left, inner sep=2pt},
  strong/.style={box, fill=black!18},
  small/.style={box, fill=black!7},
  det/.style={box},
  lab/.style={font=\footnotesize\bfseries, anchor=west}]
\node[lab] (l1) {Scaffold};
\node[strong, below=of l1.south west, anchor=north west] (a1) {Skeleton planner: structure, connectors, intent per step (1 call)};
\node[det, below=of a1] (a2) {About 20 structural normalisers};
\node[strong, below=of a2] (a3) {Structural critic (gated, optional)};
\node[lab, below=2.2mm of a3.south west, anchor=north west] (l2) {Fill};
\node[small, below=of l2.south west, anchor=north west] (b1) {Per-step fill in data-dependency waves (parallel)};
\node[det, below=of b1] (b2) {Data-flow lint $\rightarrow$ rewrite $\rightarrow$ refill of flagged steps};
\node[lab, below=2.2mm of b2.south west, anchor=north west] (l3) {Verify};
\node[det, below=of l3.south west, anchor=north west] (c1) {Platform validator + triage of messages};
\node[strong, below=of c1] (c2) {Correction routed to the step ($\leq 5$ cycles)};
\node[det, below=of c2] (c3) {Engine test: enumerated happy path, mocks, probes};
\node[strong, below=of c3] (c4) {Parameter critic and test-time repair (gated)};
\node[det, below=of c4] (c5) {Finalisation and report};
\end{tikzpicture}
\caption{Stages of the scaffolded builder. Dark grey: strong model. Light grey: small model. White: deterministic code and platform services.}
\label{fig:arch}
\end{figure}
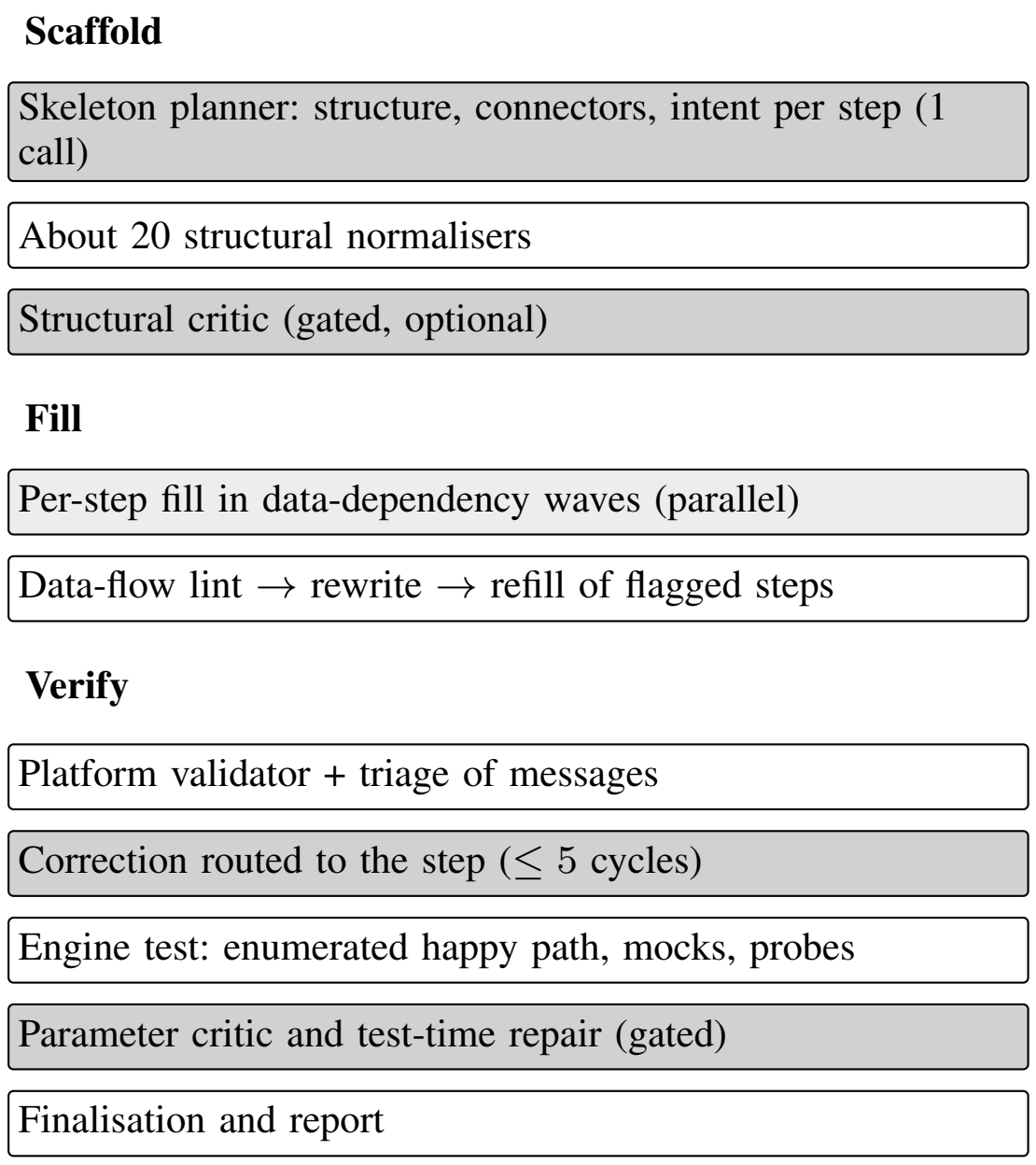

Figure~\ref{fig:arch} shows the stages; the principles that cut across them follow the three phases.

\subsection{Scaffold}

A first call to the strong model extracts from the specification and the design a requirements checklist of at most thirty items (credentials, endpoints, business rules, error handling), which the later stages receive.
One streamed call to the strong model then receives the specification, the workflow design, the connector catalogue and the building-block patterns, fetched beforehand from the catalogue the production builder also uses, and returns a skeleton: tracks, steps, branches, loops, error tracks, the connector chosen for each step, and a one-line statement of intent per step.
Parameters are left as placeholders.
About twenty deterministic normalisers then repair what the model reliably gets wrong at this level: identifiers, track references, the shape of choice, loop and parallel constructs, alias lengths and session variables.
An optional structural critic may propose changes to the skeleton; a change is kept only if it introduces no new structural defect and uses only catalogue connectors.

\subsection{Fill}

Each step's parameters are produced by a separate call to a small model.
The call sees the step's intent, the connector's schema and curated documentation, the shape of the message at that point, and the aliases of the upstream steps that are visible from it.
Steps are filled in at most three \emph{waves}.
A step waits for the transformation steps whose output it reads, by reference or as its incoming message, because their output fields exist only once they are filled; other connectors have fixed output schemas.
Steps in the same wave run in parallel.
A static data-flow analysis over the filled graph then looks for references to data that is out of scope, has the wrong shape, or was overwritten.
Some findings are rewritten deterministically; the rest trigger a refill of only the flagged steps.
Accounts and credentials are always assigned by the strong model.

\subsection{Verify}

The definition goes to the platform validator.
Every message is classified before anything is repaired:

\begin{itemize}
\item \emph{Transient}: the validator or a service behind it failed. The call is retried.
\item \emph{False positive}: the pipeline definition is valid, but the validator reports a spurious error. Table~\ref{tab:triage} lists the classes we found.
\item \emph{Provisioning}: the test environment lacks something (for example a connector is not registered there). It is not counted against the builder.
\item \emph{Genuine}: everything else.
\end{itemize}

Only genuine errors enter the correction loop.
Each is routed to the step that produced it and that step is refilled, with the error message and the signatures of the functions involved, for at most five cycles and with a stop rule when the error set no longer shrinks.
A final set of deterministic safety nets handles patterns that models keep producing regardless of instruction.

The definition is then executed in the real engine.
The harness enumerates a happy path through the control flow, stubs external connectors with responses generated for that path, and inserts \emph{probes} that record the payload at chosen points, so that a run that completes with an empty payload is not counted as a success.
A failed test triggers a repair that is kept only if the test then passes; otherwise the definition before the repair is restored.

\begin{table}[t]
\caption{Validator false-positive classes found during the study}
\label{tab:triage}
\centering
\footnotesize
\begin{tabular}{@{}p{0.50\columnwidth}@{\hspace{3mm}}p{0.40\columnwidth}@{}}
\toprule
Class & How it is recognised \\
\midrule
A reference to a platform global variable or another resolver is rejected because the validator does not register it & message pattern \\
A field name in a reference is parsed as a function call & message pattern plus the original expression \\
An expression is evaluated with empty placeholders and fails on conversion & message pattern (empty input) \\
A valid loop-iterator reference is rejected by the grammar & message pattern \\
\bottomrule
\end{tabular}
\end{table}

\subsection{Principles}

Three rules shaped most design decisions.

\textbf{Division of labour.} LLM calls are restricted to semantic and structural decisions, whereas syntactic and schema enforcement is handled deterministically.

\textbf{Local context.} After skeleton generation, each LLM prompt is scoped to a single step and its immediate data-flow dependencies rather than the entire JSON document.
The largest request in our runs stayed far below the context window, which also keeps every call in the cheaper price tier.

\textbf{Monotonic acceptance.} Every LLM change after the fill is applied to a copy of the last accepted definition, and both go through the same deterministic passes before they are compared on deterministic measures of the whole pipeline: genuine validation errors, structural defects, data-flow findings and the test outcome.
The change is kept only if no measure gets worse and no new error appears on a step it touched or on a step that had none, so a fix that breaks another step is rejected; correction stages also require that the number of errors or of failing steps falls.

\section{Evaluation design}
\label{sec:design}

The builders measure themselves differently, so comparing their own reports is biased; this section describes the instruments we built to compare them.

\subsection{Case sets}

Table~\ref{tab:data} describes the four case sets.
Three come from customer integration projects.
Engineering assembled \Done{} from recent specifications handled by the professional-services and forward-deployed engineers who implement projects for customers, varying size and complexity; product staff assembled \Dtwo{} by similar criteria, aiming at variety and higher complexity.
Both served as regression sets while the scaffolded builder was developed.
\Hreal{} was assembled after development in the same way as \Done{}, leaving out projects whose implemented pipeline had fewer than 40 steps, and was not used in the development of either builder.
\Hsyn{} contains synthetic specifications that product staff generated with LLM agents for the held-out evaluation; they cover common integration patterns, such as record synchronisation, batch extraction and delivery, enrichment, orchestration and failed-event handling.
One project appears in both \Done{} and \Dtwo{}, with different versions of its specification.
No customer data appears in this paper; case identifiers are withheld.

We characterise a case by its input, because output size depends on the builder: the length of the specification, the length of the workflow design, and the number of activities in the design.
The sets cover different ranges with overlap.
\Hsyn{} sits at the low end, with shorter specifications and fewer design activities than any customer set, and \Dtwo{} and \Hreal{} sit at the high end.
Output size correlates with the workflow design rather than with the prose specification: over the \setPoolCases{} cases the number of design activities has a Spearman correlation of \setRhoActs{} with the number of steps the scaffolded builder generates, against \setRhoSpec{} for specification length, and the builder expands the design by a median factor of \setExpansion.
The two builders produce definitions of different size from the same input, the scaffolded builder's being larger on three of the four sets.
The Campaign~B definitions of both builders use \connBAll{} distinct connectors across the \connBCases{} cases.

\begin{table}[t]
\caption{Case sets. Median [Q1--Q3] per case.}
\label{tab:data}
\centering
\footnotesize
\setlength{\tabcolsep}{3pt}
\begin{tabular}{@{}lrrrr@{}}
\toprule
 & \Done{} & \Dtwo{} & \Hreal{} & \Hsyn{} \\
\midrule
Origin & customer & customer & customer & synthetic \\
Role & develop. & develop. & held out & held out \\
Cases & 21 & 14 & 18 & 14 \\
Runs$^{a}$ & \condSkDoneRuns/\condPrDoneRuns{} (A) & \condSkDtwoRuns/\condPrDtwoRuns{} (A) & \condSkBHrealRuns/\condPrBHrealRuns{} (B) & \condSkBHsynRuns/\condPrBHsynRuns{} (B) \\
\midrule
Specification$^{b}$ & \setDoneSpec & \setDtwoSpec & \setHrealSpec & \setHsynSpec \\
Design$^{b}$ & \setDoneDesign & \setDtwoDesign & \setHrealDesign & \setHsynDesign \\
Design activities & \setDoneActs & \setDtwoActs & \setHrealActs & \setHsynActs \\
Steps, scaff.$^{c}$ & \setDoneStepsSk & \setDtwoStepsSk & \setHrealStepsSk & \setHsynStepsSk \\
Steps, baseline$^{c}$ & \setDoneStepsPr & \setDtwoStepsPr & \setHrealStepsPr & \setHsynStepsPr \\
\bottomrule
\multicolumn{5}{@{}l}{$^{a}$Scaffolded/baseline runs (campaign); \Done{}, \Dtwo{} also \condSkBDoneRuns/\condPrBDoneRuns{} in B.}\\
\multicolumn{5}{@{}l}{$^{b}$Thousands of characters.}\\
\multicolumn{5}{@{}l}{$^{c}$Steps in the final pipeline, mean over a case's runs.}
\end{tabular}
\end{table}

\subsection{Campaigns}

We ran two campaigns.
In each, the runs of the two builders were interleaved on the same days, in the same test environment.
Both builders use the same two models from one vendor: Claude Sonnet~5.5, the vendor's mid-tier model, as the \emph{strong} model and Claude Haiku~5.5 as the \emph{small} one.
The baseline also makes at most two calls per case to Gemini~3.6 Flash.
\emph{Campaign A} (late October) compares version~88 of the scaffolded builder with the production builder of that week on \Done{} and \Dtwo{}.
\emph{Campaign B} (mid-November) compares the then-current versions of both on the two held-out sets and on reruns of \Done{} and \Dtwo{}.
Campaign~A was planned as the main comparison; Campaign~B was added to test its conclusions on cases neither builder had been tuned on. We never pool the two, because both systems and the platform changed in between.

\subsection{Primary metric}

A build \emph{passes} if the final pipeline definition is valid and the engine executes it without an exception.
This is the production builder's own acceptance criterion, applied to every build: a build with no artifact, or with an artifact that was never tested, counts as a failure, since averaging only over tested builds favours whichever builder stops earlier.
Here a test counts as passed when the engine runs without an exception; the scaffolded builder's own test, which also fails a run that the engine completed although a connector failed with \code{failOnError} off or empty data reached an outbound step, failed on \skOwnStrictB{} Campaign~B builds that count as passes here.

We then addressed the remaining asymmetries.

\emph{Validity filter.} Only the scaffolded builder triaged validator messages.
We therefore revalidated offline the final artifacts of \emph{both} builders, with one validator version and one triage per campaign: 273 artifacts in late October (from all October conditions, including the earlier baseline and the critic-free mode of Section~\ref{sec:results}) and \revalBArtifacts{} in mid-November.
We report \emph{symmetric pass}: valid after triage, and executed by the engine.
The authors, whose team also maintains the validation layer, defined the triage classes from recurring validator messages while developing the scaffolded builder; two were added after a manual review of the baseline's remaining errors, before symmetric pass was computed. No independent party reviewed them. In the product, messages classified as false positives are not shown to users, except in the builder's final report.
Raw validity from the offline run matched the stored value for every baseline artifact.

\emph{Untested artifacts.} The baseline does not test what it considers invalid.
After triage, seven baseline artifacts of Campaign~A turned out valid but untested.
We executed them with the scaffolded builder's harness, three times each, and took the majority outcome: five passed and two failed, unanimously.
Campaign~B artifacts were treated in the same way. To check whether the two harnesses agree, we then ran every Campaign~B artifact that passed its builder's own test through the scaffolded builder's harness in the same way, without repair: \ctFullPrPass{} of the baseline's \ctFullPrN{} passed and all \ctFullSkN{} of the scaffolded builder's. Of the \ctFullPrFail{} baseline failures, \ctFullPrFailMock{} were business checks of the pipeline firing on mock data too thin for it; the rest were time-outs and configuration errors.

\emph{Raw floor.} We also report pass with raw validity for both builders, which ignores triage entirely and is the comparison least favourable to the scaffolded builder.

We fixed the symmetric definition after Campaign~A had run, when these asymmetries became visible.
It is less favourable to the scaffolded builder than the criterion we started with, each builder's own report of validity (\cmpAAllPassCrit{} \cmpAAllPassCritCI{} over both sets).

\subsection{Secondary metrics}

We compute \emph{cost} from per-call token logs and a price table fixed for the study, and report it relative to the baseline's mean cost per case on \Done{} in Campaign~A (1.00).
\emph{Build time} is wall-clock time per case.
An LLM judge scores each artifact once against the specification on four criteria (endpoints covered, business rules implemented, error handling, credentials), each on a 1--5 rubric that we rescale to 0--1; the judge score is their mean, and we also report the rules criterion on its own.
The judge model is GPT-5.2 (OpenAI), so a preference of a model for its own output does not explain score differences.
The rubric and scale were last revised before Campaign~A, and all scores we report use that revision.
Between the campaigns the judge prompt changed (it now receives the complete design, specification and artifact), so scores are comparable between builders within a campaign and not across campaigns; Campaign~A artifacts were scored at generation time and not re-judged.
We treat both judge scores as secondary because the judge is noisy between runs and does not detect mechanical defects.

\subsection{Statistics}

Each run of a builder over a set produces one build per case, and every case is built several times per builder (Table~\ref{tab:data}), since generation is not deterministic~\cite{ouyang2025nondeterminism}.
A case's pass rate is the share of its builds that pass, and a set's rate is the mean over its cases, so every case weighs the same whatever its number of builds; the other metrics are averaged in the same way.
Comparisons are paired by case, not by build: the builds of the two builders on a case are independent generations, and the paired difference is the mean over cases of the difference between the two builders' case means.
Confidence intervals come from a two-stage percentile bootstrap~\cite{efron1994bootstrap} (5{,}000 resamples) over cases and then builds within each case, so replicas of a case are not treated as independent~\cite{miller2024errorbars}; for the primary outcome we also report a two-sided sign-flip permutation test on per-case differences.
Relative differences in cost and time are ratios of means.
Some cases belong to the same customer project (\Dtwo{}: \setDtwoProjects{} projects; \Hreal{}: \setHrealProjects), but each is a separate pipeline with its own specification and design, so we resample cases; resampling projects instead changes interval widths in those sets by factors between \clusterWidthMin{} and \clusterWidthMax{} and changes no conclusion; for symmetric pass in Campaign~A it gives \primProjCI.
Results for several case sets, relative differences included, are combined as a mean weighted by the number of cases, with intervals from a stratified bootstrap; cases from different sets are never averaged together directly.

\subsection{Exclusions}

A build is excluded only for a documented cause outside the builder under test, and every exclusion is listed.
In Campaign~A two baseline builds in \Dtwo{} were lost to an expired cloud credential.
In both campaigns, we excluded \refusAll{} baseline builds (\refusA{} in A, \refusB{} in B) that stopped at the start because our case inputs lacked a resource-binding section that the baseline's prompt requires.
In Campaign~B two runs interrupted by service restarts, one per builder, were discarded whole, and four cases of one baseline run on \Dtwo{} were executed again after another interruption.
Builder failures, including time-budget overruns, count as failures.

\section{Results}
\label{sec:results}

\subsection{Campaign A: development sets, October}

\begin{table}[t]
\caption{Campaign A (version 88 vs.\ production builder of the same week). Per-case means over runs.}
\label{tab:campA}
\centering
\footnotesize
\begin{tabular}{@{}llrrrr@{}}
\toprule
Set & Builder & Sym.\ pass & Cost & Time & Judge \\
 & & (\%) & (rel.) & (s) & (0--1) \\
\midrule
\Done{} & scaffolded & \condSkDonePassSym & \condSkDoneCostIdx & \condSkDoneTime & \condSkDoneJudge \\
        & baseline   & \condPrDonePassSym & \condPrDoneCostIdx & \condPrDoneTime & \condPrDoneJudge \\
\Dtwo{} & scaffolded & \condSkDtwoPassSym & \condSkDtwoCostIdx & \condSkDtwoTime & \condSkDtwoJudge \\
        & baseline   & \condPrDtwoPassSym & \condPrDtwoCostIdx & \condPrDtwoTime & \condPrDtwoJudge \\
\midrule
\multicolumn{6}{@{}l}{\emph{Paired difference, scaffolded minus baseline (95\% CI), both sets}} \\
\multicolumn{2}{@{}l}{Symmetric pass} & \multicolumn{4}{l}{\cmpAAllPassSym{} \cmpAAllPassSymCI} \\
\multicolumn{2}{@{}l}{Pass, raw validity} & \multicolumn{4}{l}{\cmpAAllPassRaw{} \cmpAAllPassRawCI} \\
\multicolumn{2}{@{}l}{Cost per case} & \multicolumn{4}{l}{\cmpAAllCost{} \cmpAAllCostCI} \\
\multicolumn{2}{@{}l}{Cost per passing case} & \multicolumn{4}{l}{\cmpAAllCps{} \cmpAAllCpsCI} \\
\multicolumn{2}{@{}l}{Build time} & \multicolumn{4}{l}{\cmpAAllTime{} \cmpAAllTimeCI} \\
\multicolumn{2}{@{}l}{Judge score} & \multicolumn{4}{l}{\cmpAAllJudge{} \cmpAAllJudgeCI} \\
\multicolumn{2}{@{}l}{Rules implemented} & \multicolumn{4}{l}{\cmpAAllRules{} \cmpAAllRulesCI} \\
\bottomrule
\end{tabular}
\end{table}

Table~\ref{tab:campA} gives the results.
The scaffolded builder passed the symmetric criterion more often, \cmpAAllPassSym{} \cmpAAllPassSymCI{} over both sets.
The effect is concentrated in \Done{} (\cmpADonePassSym{} \cmpADonePassSymCI); in \Dtwo{} the interval includes zero (\cmpADtwoPassSym{} \cmpADtwoPassSymCI).
The advantage rests on \primDonePos{} of the \primDoneN{} cases of \Done{}, with none against, and the sign-flip test gives $p=\primSignFlipP$; likelihood-ratio tests in a beta-binomial model and in a logistic model with a random intercept per case agree ($p=\primBetaBinP$, $p=\primGlmmP$).

With raw validity for both builders it is \cmpAAllPassRaw{} \cmpAAllPassRawCI{}; the direction is unchanged, but the interval includes zero.
The size of the acceptance advantage thus depends on how validator false positives are treated, which is why we report both ends.
Triage adds more to the scaffolded builder's rate than to the baseline's (in \Done{}, \condSkDonePassRaw\% to \condSkDonePassSym\% against \condPrDonePassRaw\% to \condPrDonePassSym\%), as expected, since the scaffolded builder leaves false-positive messages in place, while the baseline rewrites the definition until the validator reports no message or the builder stops.

The cost and time results do not depend on that choice.
Cost per case changed by \cmpAAllCost{} \cmpAAllCostCI{} and wall-clock time by \cmpAAllTime{} \cmpAAllTimeCI{}, and the two sets agree closely.
Per passing case the cost difference is \cmpAAllCps{} \cmpAAllCpsCI.
The judge score is slightly higher (\cmpAAllJudge{} \cmpAAllJudgeCI) and the share of rules implemented is not different (\cmpAAllRules{} \cmpAAllRulesCI).

The builders also fail in different ways.
By its own criterion the baseline failed \failAPrFails{} of \failAPrBuilds{} builds, all because the definition never became valid; it does not test a definition it considers invalid, and none of its tested builds failed.
The scaffolded builder failed \failASkFails{} of \failASkBuilds{}: \failASkInvalid{} invalid definitions, \failASkTest{} test failures and 1 time-budget overrun on the largest pipeline.

\emph{Replication.} Three further runs of version~88 on \Done{} six days later, with identical configuration, passed the builder's own criterion on \replSkDonePassCrit\% of cases, against \condSkDonePassCrit\% in the campaign runs. These runs, with no same-day baseline, are not in Table~\ref{tab:campA}.

\subsection{Critic stages and cost by phase}

Running version~88 without its critic stages on \Dtwo{} cut cost by \cmpCriticDtwoCostAbs{} and time by \cmpCriticDtwoTimeAbs.
No difference in symmetric pass was detected (\cmpCriticDtwoPassSym{} \cmpCriticDtwoPassSymCI, a wide interval), but the share of business rules implemented changed by \cmpCriticDtwoRules{} \cmpCriticDtwoRulesCI{} and the judge score by \cmpCriticDtwoJudge{} \cmpCriticDtwoJudgeCI.
The critic-free mode is kept as a faster and cheaper draft option.

\emph{Cost by phase.} The builder's report records the tokens and calls of each phase.
Priced with the same table, they reproduce the recorded cost of the Campaign~B development runs within 1\% (median over builds).
On those runs the requirements and skeleton calls account for 49\% of the cost, the two critic stages for 24\%, the fill for 8\% (42 small-model calls per case on average, plus account assignment), correction, mock generation and test repair for 16\%, and other calls for 4\% (shares rounded).
The shares do not estimate the cost of a phase run alone.

\subsection{Campaign B: November}

\begin{table}[t]
\caption{Campaign B (then-current versions). Per-case means over runs.}
\label{tab:campB}
\centering
\footnotesize
\setlength{\tabcolsep}{5pt}
\begin{tabular}{@{}llrrrr@{}}
\toprule
Set & Builder & Sym.\ pass & Cost & Time & Judge \\
 & & (\%) & (rel.) & (s) & (0--1) \\
\midrule
\Done{}  & scaffolded & \condSkBDonePassSym & \condSkBDoneCostIdx & \condSkBDoneTime & \condSkBDoneJudge \\
         & baseline   & \condPrBDonePassSym & \condPrBDoneCostIdx & \condPrBDoneTime & \condPrBDoneJudge \\
\Dtwo{}  & scaffolded & \condSkBDtwoPassSym & \condSkBDtwoCostIdx & \condSkBDtwoTime & \condSkBDtwoJudge \\
         & baseline   & \condPrBDtwoPassSym & \condPrBDtwoCostIdx & \condPrBDtwoTime & \condPrBDtwoJudge \\
\Hreal{} & scaffolded & \condSkBHrealPassSym & \estSkBHrealCostIdx$^{e}$ & \condSkBHrealTime & \condSkBHrealJudge \\
         & baseline   & \condPrBHrealPassSym & \estPrBHrealCostIdx$^{e}$ & \condPrBHrealTime & \condPrBHrealJudge \\
\Hsyn{}  & scaffolded & \condSkBHsynPassSym & \estSkBHsynCostIdx$^{e}$ & \condSkBHsynTime & \condSkBHsynJudge \\
         & baseline   & \condPrBHsynPassSym & \estPrBHsynCostIdx$^{e}$ & \condPrBHsynTime & \condPrBHsynJudge \\
\midrule
\multicolumn{6}{@{}l}{\emph{Paired difference, scaffolded minus baseline (95\% CI)}} \\
\multicolumn{2}{@{}l}{Sym.\ pass, development} & \multicolumn{4}{l}{\cmpBDevPassSym{} \cmpBDevPassSymCI} \\
\multicolumn{2}{@{}l}{Sym.\ pass, held-out} & \multicolumn{4}{l}{\cmpBAllPassSym{} \cmpBAllPassSymCI} \\
\multicolumn{2}{@{}l}{Cost per case, development} & \multicolumn{4}{l}{\cmpBDevCost{} \cmpBDevCostCI} \\
\multicolumn{2}{@{}l}{Build time, development} & \multicolumn{4}{l}{\cmpBDevTime{} \cmpBDevTimeCI} \\
\multicolumn{2}{@{}l}{Build time, held-out} & \multicolumn{4}{l}{\cmpBAllTime{} \cmpBAllTimeCI} \\
\multicolumn{2}{@{}l}{Judge score, development} & \multicolumn{4}{l}{\cmpBDevJudge{} \cmpBDevJudgeCI} \\
\multicolumn{2}{@{}l}{Judge score, held-out} & \multicolumn{4}{l}{\cmpBAllJudge{} \cmpBAllJudgeCI} \\
\bottomrule
\multicolumn{6}{@{}l}{$^{e}$Estimated from token counts for runs without per-call events.}
\end{tabular}
\end{table}

Table~\ref{tab:campB} gives the results three weeks later under the same symmetric criterion.
Both builders pass almost every case, so the criterion is near its ceiling, and we detected no difference: \cmpBDevPassSym{} \cmpBDevPassSymCI{} over the development sets and \cmpBAllPassSym{} \cmpBAllPassSymCI{} over the held-out sets, where the baseline passed every case, and the intervals bound any difference.
On the held-out sets, triage decides whether a difference is detected.
There it accepted \triBSkHo{} scaffolded artifacts and \triBPrHo{} baseline artifact, from \triBHoCases{} cases, that raw validation rejected, all for one reason: a reference to a platform global variable, which the validator does not resolve (first class in Table~\ref{tab:triage}).
All \triBHoN{} executed in the engine, the baseline's in the cross-test, since the baseline had not tested it; over all Campaign~B sets, \triBExec{} of the \triBN{} artifacts admitted by triage did.
With raw validity for both builders the estimates are \cmpBDevPassSymRawOff{} \cmpBDevPassSymRawOffCI{} and \cmpBAllPassSymRawOff{} \cmpBAllPassSymRawOffCI, the latter favouring the baseline.
The test harness moves the estimate the other way.
Judging both builders by the scaffolded builder's harness gives \cmpBDevPassSymCommon{} \cmpBDevPassSymCommonCI{} on the development sets and \cmpBAllPassSymCommon{} \cmpBAllPassSymCommonCI{} on the held-out sets, and \cmpBAllPassSymCommonNM{} \cmpBAllPassSymCommonNMCI{} on the held-out sets if failures on too-thin mocks are not counted.
On held-out cases, then, the sign depends on the instrument, and we report no acceptance difference.

The efficiency advantage is larger than in Campaign~A.
On the development sets, where every run recorded per-call events, cost per case changed by \cmpBDevCost{} \cmpBDevCostCI{} and build time by \cmpBDevTime{} \cmpBDevTimeCI.
On the held-out sets build time changed by \cmpBAllTime{} \cmpBAllTimeCI.
Because of a defect in our event store, most held-out runs lack per-call events, so their cost is estimated: for each builder, a linear model of cost on four token classes, fitted on the builds of the same version that have events (\costSkBEvCases{} and \costPrBEvCases), predicts the mean cost of a left-out run within \estSkBLoroErr\% and \estPrBLoroErr\%.
The estimates put the scaffolded builder's cost \estBHrealCostRed{} lower on \Hreal{} and \estBHsynCostRed{} lower on \Hsyn{}; we give them without intervals.
The judge score is higher for the scaffolded builder on the development sets and on the held-out sets, and the share of rules implemented is not different (\cmpBDevRules{} \cmpBDevRulesCI{} and \cmpBAllRules{} \cmpBAllRulesCI).
Over all four sets, the criterion that differs most is credentials (\judgeBCred{} \judgeBCredCI), where the baseline also lacked its resource section (Section~\ref{sec:threats}); error handling does not differ (\judgeBErr{} \judgeBErrCI).

\subsection{Comparison of the two campaigns}

The reruns of the development sets indicate why the acceptance gap of Campaign~A did not reappear in Campaign~B.
On \Done{}, the baseline's symmetric pass rose from \condPrDonePassSym\% in October to \condPrBDonePassSym\% in November, while the scaffolded builder's changed little (\condSkDonePassSym\% and \condSkBDonePassSym\%).
On \Dtwo{}, where the October interval already included zero, both changed little (baseline \condPrDtwoPassSym\% and \condPrBDtwoPassSym\%; scaffolded \condSkDtwoPassSym\% and \condSkBDtwoPassSym\%).
The convergence is therefore not explained by a difference in difficulty between held-out and development sets.
It is consistent with changes in the production baseline, in particular its larger correction budget, but the test environment changed in the same period and our data do not separate the two.
Over the same period the baseline's cost per case grew by factors of \prDoneCostGrowth{} on \Done{} and \prDtwoCostGrowth{} on \Dtwo{}, its build time by \prDoneTimeGrowth{} and \prDtwoTimeGrowth, and its calls to the small model, which serves validation and reporting, from \prDoneSmallCallsA{} to \prDoneSmallCallsB{} per case on \Done{}.

Taken together, the two campaigns support two claims.
On acceptance, under the symmetric criterion no case set in either campaign showed a difference against the scaffolded builder with an interval that excludes zero, and only \Done{} in October showed one in its favour.
On efficiency, it cost less and built faster in every comparison.

\section{Lessons learned}
\label{sec:lessons}

\subsection{Architecture}

These lessons come from the development history of the scaffolded builder and were not measured in the campaigns.

\textbf{Deterministic post-processing fixed recurring structural defects.}
Recurring structural or syntactic defects that we addressed with an instruction returned or moved elsewhere; the same defects handled by a normaliser or a final safety net did not recur.

\textbf{Dependency-ordered generation reduces data-flow reference errors.}
Data-flow defects in our development runs dropped markedly once steps were filled in dependency order, so that each prompt receives the concrete output schemas of upstream steps rather than descriptions of them.

\textbf{Small models are effective when constrained to localised, schema-bound tasks.}
In development runs, moving the per-step fill to the small model cut cost substantially without a loss of pass rate that we could detect.
Moving correction or account assignment to it raised builder failures, mostly on nested function expressions.
Price per token did not predict cost per task: reasoning tokens billed as output, prompt caching and price tiers by request size dominated.

\subsection{Evaluation}

\textbf{Validator errors affect both the builder and its evaluation.}
The platform validator rejected correct definitions in the ways listed in Table~\ref{tab:triage} and accepted incorrect ones: a numeric error code written as a decimal passed validation and was rejected by the engine only on the path where that step executed.
During development, repairing every validator message damaged correct steps.
In the campaigns, scoring by raw validity penalised whichever builder triggered more false positives, and in Campaign~B a single false-positive class decided whether the held-out comparison showed a difference (Section~\ref{sec:results}).
Acceptance should therefore be decided by execution, with the same triage applied to every system compared.

\textbf{A stop condition that fails to trigger can double the cost of an agent loop.}
In an earlier version of the baseline, the stop condition after a successful test did not trigger, and the loop kept validating and testing until its token budget ran out.
When we fixed it, its cost and time roughly halved with no change in pass rate; against that earlier version the scaffolded builder's cost and time on \Done{} had been \cmpOldDoneCostAbs{} and \cmpOldDoneTimeAbs{} lower, a comparison that would have overstated the architecture's contribution.

\textbf{The production baseline changed during the study.}
Its main change between the campaigns was a larger number of correction cycles, made to raise acceptance; the rise in its small-model calls (Section~\ref{sec:results}) is consistent with this, but we did not isolate its effect.
A comparison against a live system should state the baseline's version and date and measure both sides on the same days.

\textbf{Measurement instruments also contained defects.}
A large share of our effort went into instruments that were wrong: a mean computed only over tested cases; an optional review stage that failed silently and was reported as having run; and our own offline revalidation, which reported false ordering errors after a normalisation renamed the root track.
Each was found by reconciling independent sources (per-call events, the builder's report and the tool trace).

\section{Threats to validity}
\label{sec:threats}

\emph{Construct.} Passing means valid and executing on a happy path with mocked services; it does not mean the pipeline does what the customer needs.
The triage classes were defined by the authors, whose team maintains the validator, and in Campaign~B they decide whether an acceptance difference is detected on held-out cases, through a single class whose artifacts all executed in the engine.
The judge was not validated against independent human ratings, so its scores are a weak, secondary proxy.

\emph{Internal.} The authors developed both builders, which removes the risk of a weakly configured baseline but not that of unconscious preference; the protocol of Section~\ref{sec:design} is intended to limit it.
Apart from the critic stages, we did not ablate individual design decisions, so the differences between the builders cannot be attributed to any one of them.
Our case inputs lacked the resource section that the baseline's prompt treats as its only source of accounts and global variables; most baseline builds proceeded without it, which may lower its credentials scores.
Both builders call the same validator and test engine, but each prepares its own test (input, mocked responses and checks); all passing Campaign~B artifacts also went through the scaffolded builder's test, which may suit the definitions of the builder it was written for, and the reverse is not possible, because the baseline's test is part of its agent loop.
The data-flow analysis is used inside the scaffolded builder, so we do not use its findings as an outcome.
The test environment changed between campaigns (for instance, a connector was registered that had been missing), so results from different dates are not directly comparable.

\emph{External.} The study covers a single platform and DSL, one main model family, and 67 cases.
The customer cases were taken from recent projects with a check on size only (\Dtwo{} also aimed at higher complexity), but they are not a formal random sample of customer demand.
\Hsyn{} was generated with the builders' strong model (Claude Sonnet~5.5) and may be easier for both.

\emph{Conclusion.} Two to four runs per condition separate only large differences in pass rate; intervals for \Dtwo{} and \Hreal{} are wide.
We made several comparisons without correction for multiplicity. Because the symmetric criterion was fixed after Campaign~A ran, its acceptance result is exploratory; Campaign~B tested it and did not replicate it, while the efficiency result held in both.

\emph{Artifacts.} The specifications, workflow designs and pipeline definitions, including those of \Hsyn{}, use proprietary formats and are not released.
Sections~\ref{sec:arch}--\ref{sec:design} and Table~\ref{tab:triage} describe the architecture, the protocol and the false-positive classes in enough detail to apply them on another platform.

\section{Related work}
\label{sec:related}

\emph{Low-code workflow generation in industry.}
The closest work generates workflows for an enterprise platform by first producing the outline and then populating each step~\cite{ayala2025lowcode}, with companion papers on retrieval to reduce hallucinated steps~\cite{bechard2024hallucination} and on fine-tuned small models~\cite{ayala2025slm}.
WorkTeam separates planning, component arrangement and parameter filling across agents~\cite{liu2025workteam}, and classifier-augmented generation predicts the stages of data-integration workflows and then their edges and properties~\cite{gschwind2025cag}.
Work on another DSL for automation workflows compares fine-tuning with retrieval~\cite{bassamzadeh2024dsl}.
These systems are evaluated by similarity to reference workflows or by structural error rates; none reports execution, a loop that feeds validator errors back to the generator, or a paired comparison with a production predecessor.

\emph{Benchmarks.}
WorkflowLLM evaluates statically~\cite{fan2025workflowllm}.
Recent benchmarks do execute generated workflows~\cite{mcpgen2026,chat2workflow2026,xue2025comfybench}, in academic settings and with little or no repair loop.
Prompt2DAG finds that structure-first strategies are far more reliable than direct generation for data pipelines~\cite{prompt2dag2025}.

\emph{Decomposition and repair.}
Generating a skeleton and expanding it in parallel~\cite{ning2024sot}, decoding a sketch before details~\cite{dong2018coarse} and planning before coding~\cite{jiang2024selfplanning} are established.
Repair driven by execution feedback is also established~\cite{chen2024selfdebug}, although its benefit shrinks when cost is counted~\cite{olausson2024selfrepair}.
Grammar prompting~\cite{wang2023grammar} and constrained decoding for JSON Schema~\cite{geng2025jsonschemabench} enforce syntactic or schema validity, not data flow or runtime semantics.

\emph{Industrial reports.}
Industrial reports on LLM tools typically give acceptance or success rates on internal data without a controlled baseline~\cite{alshahwan2024testgen,rondon2025passerine}.
In this workshop's first edition, a multi-agent system with intent-understanding, planning, execution and verification phases was compared with the retrieval baseline of a production search system~\cite{avataragent2026}.

\section{Conclusion}

Replacing a monolithic agent loop with an agentic scaffold, fill and verify design gave the platform a builder that uses fewer resources than the production builder: \cmpAAllCostAbs{} lower cost and \cmpAAllTimeAbs{} less build time in October, and \cmpBDevCostAbs{} and \cmpBDevTimeAbs{} on the same development cases three weeks later, after the production builder had raised its acceptance at a higher cost.
On acceptance the evidence is weaker: the October advantage under a criterion that triages validator false positives for both builders (\cmpAAllPassSym) is exploratory and was not detected in November, with both builders near the ceiling; on held-out cases the sign depended on the instrument.
Before release, twelve internal and customer users piloted it for three weeks and reported corner cases in three bug tickets, which were fixed.
On 16 November 2026 it became the builder offered to customers, with the former production builder kept as a fallback; we do not yet measure how often users fall back to it.
The parts we expect to transfer to other proprietary DSLs are deterministic normalisers for recurring structural errors, the triage of validator messages before repair, execution with probes as the acceptance test, monotonic gates around every model step, and a symmetric measurement protocol.

\section*{Acknowledgment}
The authors acknowledge the use of Claude (Anthropic) to assist in drafting initial text from their experimental design, data, statistical analyses, system architecture, technical outlines, and core literature references. The authors subsequently rewrote, verified, and refined all sections and take full responsibility for the final content.


\end{document}